\documentclass[12pt]{article}
\usepackage{graphicx,lscape,color,url,eurosym}
\usepackage[a4paper, total={6.5in, 8in}]{geometry}
\usepackage[square]{natbib}

\title{Trends in crypto-currencies and blockchain technologies: A monetary theory and regulation perspective}

\author{Gareth W. Peters$\ddag$ $\star$ $\ast$ \and Efstathios Panayi$\dag$ $\ast$ \and Ariane Chapelle$\dag$\\
{\small{
$\ddag$ Department of Statistical Science, University College London}} \\
{\small{
$\star$ Associate Fellow, Oxford Mann Institute, Oxford University}}\\
{\small{
$\ast$ Associate Fellow, Systemic Risk Center, London School of Economics.}}\\
{\small{
$\dag$ UCL, Department of Computer Science, WC1E 6BT, London, UK}}\\
}

\begin{document}
\maketitle

\begin{abstract}

The internet era has generated a requirement for low cost, anonymous and rapidly verifiable transactions to be used for online barter, and fast settling money have emerged as a consequence. For the most part, e-money has fulfilled this role, but the last few years have seen two new types of money emerge. Centralised virtual currencies, usually for the purpose of transacting in social and gaming economies, and crypto-currencies, which aim to eliminate the need for financial intermediaries by offering direct peer-to-peer online payments. 

We describe the historical context which led to the development of these currencies and some modern and recent trends in their uptake, in terms of both usage in the real economy and as investment products. As these currencies are purely digital constructs, with no government or local authority backing, we then discuss them in the context of monetary theory, in order to determine how they may be have value under each. Finally, we provide an overview of the state of regulatory readiness in terms of dealing with transactions in these currencies in various regions of the world.  

\end{abstract}

\section{Introduction}
It has been 20 years since Bill Gates opined: `Banking is essential, banks are not'. The early 21st century has seen a proliferation of fintech (financial technology) firms, providing a wide and varied array of services, from payments and local and international money transmission through to financing through peer-to-peer lending and crowdfunding. Venture capital funding in the UK for fintech related business has increased to over \$500 million in 2014, while the sector is estimated to contribute more than GBP 20 billion to the economy\footnote{Investment Trends in Fintech report by SVB, available at \url{http://www.svb.com/News/Company-News/2015-Fintech-Report--Investment-Trends-in-Fintech/?site=uk}}. A number of countries have stated their intention to create an eco-system in which such businesses can grow, which can only mean the continued growth of the sector in the foreseeable future. 

In parallel to these innovations, which aim at reducing the friction of making payments and transfers in fiat currency, which have been facilitated by electronic money (``e-money''), there has also been a rise in the use of virtual and crypto-currencies. While the former have traditionally been utilised in virtual economies, such as those of an online game or community \citep{lehdonvirta2014virtual}, the latter has entered into the real economy also, see discussion in \citet{peters2014opening}. The goal of the most successful crypto-currency thus far, Bitcoin, is in fact in line with that of the companies mentioned above, i.e. reducing transaction costs, but with the additional aim of completely eliminating the need for financial intermediaries. 

\begin{figure}[]
	\begin{center}
\includegraphics[width=0.99\textwidth]{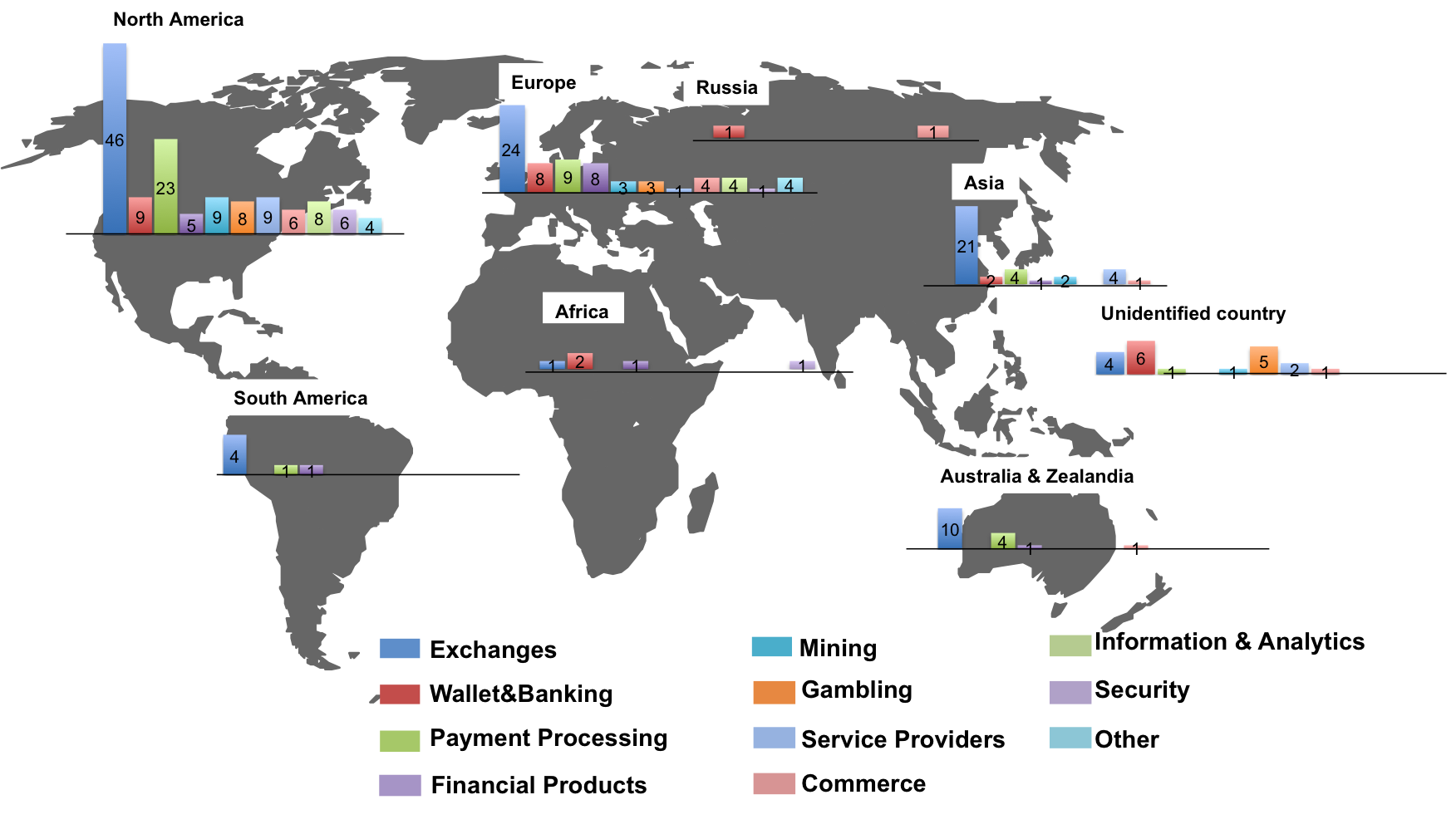}
	\label{fig:comps}
	\end{center}
	\caption{Location and industry for 318 startups in Bitcoin. Source: \protect\url{http://www.creandum.com/318-records-and-counting-the-bitcoin-database-is-now-available-for-everyone-2/}}
\end{figure}

While one of the objectives of Bitcoin was to become a form of electronic cash for online payments, its main use thus far has been for speculation. However, this is beginning to change, and there are numerous emerging intermediaries that are beginning to operate within the Bitcoin network, which include exchanges, merchant processes and money transmitters. In fact, Bitcoin has been traded in various exchanges since at least 2010\footnote{Mt. Gox was launched in July 2010, and was responsible for the vast majority of Bitcoin trading until 2013.}, and it has experienced various boom-bust cycles in this time with regard to its exchange to the US dollar, UK pound, Euro and other important fiat currencies. This price volatility is seen as an impediment to its more widespread use as a medium of exchange, and there have already been suggestions (e.g. by \citet{brito2014bitcoin}) for the creation of financial instruments to aid in the reduction of volatility. Section \ref{sec:trends} will highlight trends in price and trading volumes for Bitcoin over the past two years. 

The main innovation of crypto-currencies such as Bitcoin has been introducing technologies such as the blockchain, a ledger containing all transactions for every single unit of currency. It differs from existing ledgers in that it is decentralised, i.e. there is no central authority verifying the validity of transactions. Instead, it employs verification based on cryptographic proof, where various members of the network verify `blocks' of transactions approximately every 10 minutes. The incentive for this is compensation in the form of newly `minted' Bitcoins for the first member to provide the verification. The distributed ledger at the heart of the network could, of course be used for a number of other use-cases, such as smart property and smart contracts, and regulators have looked at such applications much more favourably than crypto-currencies, though this is also beginning to change. We provide more details of such use-cases and the potential of the blockchain in Section \ref{sec:blockchain}. 

Bitcoin in particular has had a fair amount of criticism questioning why its digital tokens, produced as a result of solving a computational problem, should have any value, particularly when they are not backed by any authority i.e. not fiat currencies. In Section \ref{sec:value} we discuss this question in more detail from both the traditional metalist views on currency value generation and more recent (and perhaps less orthodox) monetary theories, such as the Modern Monetary Theory (MMT) in this context. We discuss issues relating to monetary theory and resultant economic policy implications that may arise under each of these frameworks, if crypto-currencies were to interact more widely with the real economy.

In this environment of fast-paced technological evolution, financial innovation is running ahead of regulation. For example, the transaction anonymity provided by transacting in the Bitcoin network is a clear driver for several operational risks, money laundering, fraud and legal risk, as discussed at length in \citet{peters2014opening}.  Government responses have been mixed, and while they want to be careful not to overburden the budding sector of financial innovation with excessive regulation and curtail growth in the area, there is a need to ensure that the new services are not used to circumvent regulation in traditional banking services.  Section \ref{sec:regulation} will summarise regulatory interventions in some major economies. 

\section{Physical and electronic forms of money, and the development of crypto-currencies}
\label{sec:differences}
In this section we provide a brief overview of the historical context in which crypto-currencies have emerged. We touch upon government-backed and commodity backed currency and discuss the development of cryptographic protocols that enabled e-money. Finally, we describe the online communities which were first exposed to virtual currency and the differences between the aforementioned forms of money and crypto-currency. 

\subsection{Fiat currency and e-money}
We start with a brief definition of a fiat currency. The European Central Bank defines fiat currency as any legal tender designated and issued by a central authority that people are willing to accept in exchange for goods and services because it is backed by regulation, and because they trust this central authority. Fiat money is similar to commodity backed money in this regard with respect to its usage, but differs in that it cannot be redeemed for a commodity, such as gold. The most common form of fiat currency backing is at the sovereign state's government level, but there have also been localised currencies or private monies, see discussion in \cite{peters2014opening} for their use in local communities in the UK and Germany. 

While one is most commonly accustomed to thinking about money in its physical form, only a very small fraction of a country's total money supply is typically in the form of notes and coins. In the UK, this percentage is 2.1\% of the 2.2 trillion GBP total money supply \citep{richard2011economics}. This then motivates the discussion of electronic money, or e-money, defined by \cite{al2009development} as a floating claim on a private bank or other financial institution that is not linked to any particular account. Under this rather general definition one can consider many different forms of e-money such as bank deposits, electronic fund transfers, direct deposits, and payment processors (including micro-payments). 

Instead we put forward the rather more narrow definition of the UK regulator defines electronic money as follows (see \citet{halpin2009developments}): 
\begin{center}
``\textsl{Electronic money (e-money) is electronically (including magnetically) stored monetary value, represented by a claim on the issuer, which is issued on receipt of funds for the purpose of making payment transactions, and which is accepted by a person other than the electronic money issuer. Types of e-money include pre-paid cards and electronic pre-paid accounts for use online.}''
\end{center}
Typically, e-money is stored in the same unit of account as the fiat denomination used to obtain the e-money. 

\subsection{Cryptographically secure e-money}
In the case of early forms of e-money one may go back to the early 1980's where David Chaum (see \citet{chaum1988blind,chaum1985cryptographic,chaum1992achieving}) developed the concept of electronic cash under the view that for it to be useable in the real world economy it would require a token of money that would emulate physical currency, and most importantly, privacy feature to enable safely and securely anonymous payments. He developed such a digital cash as an extension to the RSA encryption protocol used for most security purposes on the web at present which led to the creation of the company DigiCash. Due to complications that arose with the central bank in Amsterdam where DigiCash was founded, it was decided that such currency would only be sold as a product to banks. This e-money attempt had a lot of promise, but it was unable to gain mainstream uptake in the end due more to political and business related issues \footnote{\url{http://globalcryptonews.com/before-bitcoin-the-rise-and-fall-of-digicash/}}. 

Following DigiCash there was an explosion of small venture capital firms established to develop e-money systems, leading to the release of a key initial regulatory response to such e-money, the 1994 EU Report by the Working Group on EU Payment Systems which was made to the council of the European Monetary Institute. After the release of this report there were three notable front runners that emerged: PayPal, Liberty Reserve and E-gold which was incidentally started by Nick Szabo, a former DigiCash employee and e-contract innovator.

Whilst PayPal was careful to negotiate and avoid the challenges faced by integrating into the monetary system in a manner deemed acceptable by central banks and regulators, the other two eventually ran foul of authorities in the US due to the the suspected nature of some clients that may have taken up these services for activities related to money laundering and criminal enterprise. These three early e-money systems primarily operated as centralized systems. 

The impact of e-money on physical forms of currency has been discussed by \citet{drehmann2002challenges}, while \citet{sifers1996regulating} discusses policy concerns and regulatory issues. We will now be focusing on other electronic forms of money, which in contrast to e-money are not digital representations of fiat money, but rather new forms of currency altogether.  

\subsection{Virtual currencies to facilitate online gaming economies}
The 1990s saw the emergence of virtual currencies, typically currencies that were also centralized but restricted, at least in their early forms, to use in online messaging and virtual gaming environments. An early example was the Q coin, which could be purchased from brick and mortar shops in China for use on Tencent's online messaging platform \cite{lehdonvirta2014virtual}. Virtual currencies are now prevalent in massively multiplayer online games (e.g. World of Warcraft) or life simulation games (e.g. Second Life). 

Where these currencies are used as the medium of exchange in an online virtual economy, they have similarities with their fiat currency counterparts. To start with, the currencies are typically used by the participants in the economy for the purchase of virtual goods and services. Secondly, the currencies feature a central authority, which similar to a country's central bank\footnote{The Money Supply, New York Federal Reserve, accessed 10 August 2015, available at \url{http://www.newyorkfed.org/aboutthefed/fedpoint/fed49.html}}, can regulate the money supply in order to attain particular goals, such as controlling inflation or promoting economic growth. In particular, some platforms actively manage the monetary supply, increasing money supply through in game features, or reducing  money supply through in game ``sinks'', or desirable consumption items that remove money from the online environment \citet{lehdonvirta2014virtual}.

The limited interaction of virtual currencies with the real economy stems from the fact that for many of these virtual currencies, the flows between fiat and the virtual currency are uni-directional, i.e. one can only purchase, but not sell the virtual currency \citep{peters2014opening}. For some environments, such as World of Warcraft, the developer Blizzard Entertainment actively monitors and polices the use of their virtual currency to restrict its use within the virtual economy and thus avoid any legal issues that may arise. There are a minority of cases, however, such as Second Life, whose developer Linden Labs does not oppose actively the exchange of the Linden dollar with real fiat currency. This has led to a bi-directional cross over between the virtual currency and real fiat currencies.    

Virtual currencies cannot be fully considered as e-money since, as although they share some of its attributes, there is currently no legal founding to enforce the link between fiat physical money and virtual currencies as there is in regulated electronic money transactions. In addition, virtual currencies are not stored in the same unit of account as any fiat currency that would preserve their worth. 

\subsection{Crypto-currencies}
Unlike such virtual currencies which are centrally controlled by a game designer or online platform operator, the development of crypto-currencies has been such that they are typically not operated in a centralized manner. By far the most widely known crypto-currency is Bitcoin, introduced by \cite{nakamoto2008bitcoin}. It is a `decentralized' currency, in that one does not need financial intermediaries in order to perform electronic transactions and it does not have a central bank or other authority in control of monetary policy. 

Simply put, Bitcoin can be described as a decentralised ledger of transactions. The role of the verifying third party found in centralised systems is played by the Bitcoin network participants, who contribute computational power and are rewarded in the form of new amounts of crypto-currency. Designed to be a currency for the internet, Bitcoin is not localized to a particular region or country, nor is it intended for use in a particular virtual economy. It is not backed by any local government or private organisation and is being circulated in the real economy on an increasing scale. Because of its decentralized nature, this circulation is largely beyond the reach of direct regulation, monetary policy, oversight and money supply control that has traditionally been enforced in some manner with localized private monies and e-money. 

Bitcoin is certainly not the only crypto-currency, and there are numerous papers discussing both identified weaknesses of the current protocol, as well as possible improvements to both centralised and decentralised currency architectures, see discussions in \cite{eyal2014majority,barber2012bitter,carroll2015creating} and references therein. Other examples of decentralized crypto-currencies include Litecoin, which was originally based on the Bitcoin protocol, and has a faster verification time, Ripple which is a monetary system based on trust networks, Dogecoin, Monero and Nxt.

\subsection{The distinct nature of crypto-currencies}
To distinguish between centralized and decentralized currencies, one can consider for instance the definition from the central bank of Canada \footnote{\url{http://www.bankofcanada.ca/wp-content/uploads/2014/04/Decentralize-E-Money.pdf}}
`Decentralized e-money is stored and flows through a peer-to-peer computer network that directly links users, much like a chat room. No single user controls the network.'

The ECB report on virtual currencies\footnote{\url{https://www.ecb.europa.eu/pub/pdf/other/virtualcurrencyschemesen.pdf?fe92070cdf17668c02846440e457dfd0}} classified these currencies based on their interaction with fiat money and the real economy. \cite{peters2014opening} proposed to extend this classification to include the existence of a central repository and a single administrator, where the absence of both means that the currency is operated via a decentralised network consensus-type administration. Decentralised virtual currencies are then termed crypto-currencies, as the operation of these currencies is usually based on cryptographic proof provided by a network, rather than the existence of a trusted third part to verify transactions. 

Differentiating between the different forms of virtual currencies is non-trivial as they are multifaceted in their attributes and interactions in the real economy. Several differences between centralised virtual currencies and crypto-currencies were identified in \cite{peters2014opening} and we briefly summarise some of these below:
\begin{itemize}
\item In terms of changes to their specification. In centralised virtual currencies the specification can be altered by the controlling company, whereas in crypto-currencies the specification is agreed by cryptographic consensus.  
\item In terms of their purpose and geographic area of operation, i.e. for use within an online community in the case of centralised virtual currencies, or in the wider economy, in the case of crypto-currencies. 
\item In terms of the existence of a centralised authority to exert control over issuance, monetary policy and administration of currency balances. In centralised virtual currencies, a central authority can step in to control money supply and reverse transactions at will. In crypto-currencies, the absence of a centralised authority means that users control these aspects according to the computational power they contribute to the network. In addition, transactions are generally irreversible, as there is no authority to appeal to.
\item In terms of the flow of currency between users and the currencies' exchangeability with fiat. 
\item In terms of the value generation mechanism, which will be discussed in detail in Section \ref{sec:value}.
\end{itemize} 

The distinct nature of crypto-currency is apparent in its comparison to centralised virtual currency above, but also, as we will see here, to e-money. The issuance mechanism in Bitcoin is fixed, with the coin generation process and final available currency dictated by a mathematical protocol. E-money is intrinsically linked to the underlying fiat currency, whose issuance is controlled by a central banking authority. In addition, in the current absence of the requirements of `know your customer' that e-money transactions tend to require, one can have a more anonymous interaction with crypto-currency. In general it is acknowledged that anonymity is perhaps greater with crypto-currencies, as not all companies directly follow the Financial Action Task Force standards with regard to customer identification. 

Another key point that can distinguish the utility of crypto and virtual currencies relates to the environments they operate in. This is becoming an important feature in terms of accessibility, at present Bitcoin is limited to people with internet connections. This turns out to be significant as it precludes its widespread uptake in the third world and developing countries, where e-money has been very popular in mobile and paging service networks. 

To conclude this section on the distinct nature of crypto-currency, we also observe the comments made by \cite{maurer2013perhaps} that in the case of Bitcoin, it is its code that is its core. They state succinctly: ``...the currency functions based on the trust its community of users place in the code and, as with all free and open-source projects, the trust they place in their collective ability to review, effectively evaluate, and agree as a group to changes to it''. This is clearly different from e-money which involves trust in the central authority, government or state that backs the fiat denomination underlying the e-money.

\subsection{Fulfilling the functions of money}
Having described the historical context in which crypto-currencies emerged, as well as the differences with other forms of electronic money, we now analyse whether these currencies can fulfill the traditional role of money in an economy. A widely held view is that money should serve three distinct functions:
\begin{enumerate}
\item It should be generally accepted as a medium of exchange.;
\item It should be a unit of account so that we can compare the costs of goods and services over time and between merchants.; and
\item It should be a store of value that stays stable over time.
\end{enumerate}

Both the Bank of England\footnote{\url{http://www.bankofengland.co.uk/publications/Documents/quarterlybulletin/2014/qb14q302.pdf}} and the central bank of Canada\footnote{\url{http://www.bankofcanada.ca/wp-content/uploads/2014/04/Decentralize-E-Money.pdf}}, using Bitcoin as a case study, found that crypto-currencies do not currently fulfill these functions in the way that fiat currencies and e-money do. However, it is of course possible that in the future, a more widespread uptake in a particular crypto-currency may lead it to it satisfying this criteria. This is not necessarily the view held in all jurisdictions throughout the world, we will discuss recent changes proposed to this view in for instance Australia, in Section \ref{sec:regulation}.

Separate from the \textit{functions} of money, one can also explore particular \textit{qualities} of money that make it suitable for facilitating transactions. In the case of commodity money, these include durability, value per weight unit (portability), and scarcity, and \cite{graf2015commodity} argues that Bitcoin evaluates well on each characteristic. As these currencies were primarily oriented towards direct, online transactions, we can additionally consider the following qualities in the context, e.g. of online commerce \citep{drehmann2002challenges}:
\begin{itemize}
\item They should be low cost; 
\item they should provide reliable security; and 
\item they should offer a degree of privacy in transactions. 
\end{itemize}
See further discussions on these points in \cite{maurer2013perhaps}.

Two further distinctive feature of crypto-currency like Bitcoin, which are not readily replicated in fiat e-money, relate to its divisibility and fungibility, see discussion in \cite{barber2012bitter}. They note that one of the key practical appeals of for instance Bitcoin is ``...the ease with which coins can be both divided and recombined to create essentially any denomination possible. This is an Achilles’ heel of (strongly anonymous) e-cash systems, because denominations had to be standardized to be unlinkable, which incidentally makes the computational cost of e-cash transactions linear in the amount. In Bitcoin, linkage is inherent, as it is what prevents double spending; but it is the identities that are “anonymous”.''

We note that such crypto-currencies as Bitcoin do not however have, compared to conventional fiat backed e-money payment systems, a strict governance structure other than its underlying software. The implications of this are discussed recently by both \citet{peters2014opening} and \citet{bohme2015bitcoin}. Without the lack of governance afforded by traditional fiat e-money payment systems, the Bitcoin network is unable to impose any obligation on a financial institution, payment processor, or other intermediary to verify a user’s identity or cross-check with watch-lists or embargoed countries. The implications of this for money laundering and money transmitter regulations are discussed in \citet{brito2014bitcoin}. Finally, it is clear that without central governance, one cannot impose any form of prohibition on sales of particular items, this point is discussed by \citet{maccarthy2010payment}, where they point out that traditional e-money and credit card payment systems regularly monitor and disallow a range of transactions which are deemed unlawful in the place of sale.

\section{Trends in the usage of crypto-currencies in the economy}
\label{sec:trends}
The discussion in the previous section should highlight the much greater potential of crypto-currencies for entering the real economy, compared to virtual currencies. We present in this section summary statistics for the uptake of Bitcoin, the most popular crypto-currency. We also discuss associated investment products, as well as views about the currency's potential use for facilitating criminal transactions. 

\subsection{Bitcoin trading by exchange and currency}
\begin{figure}[ht!]
	\begin{center}
\includegraphics[width=0.9\textwidth]{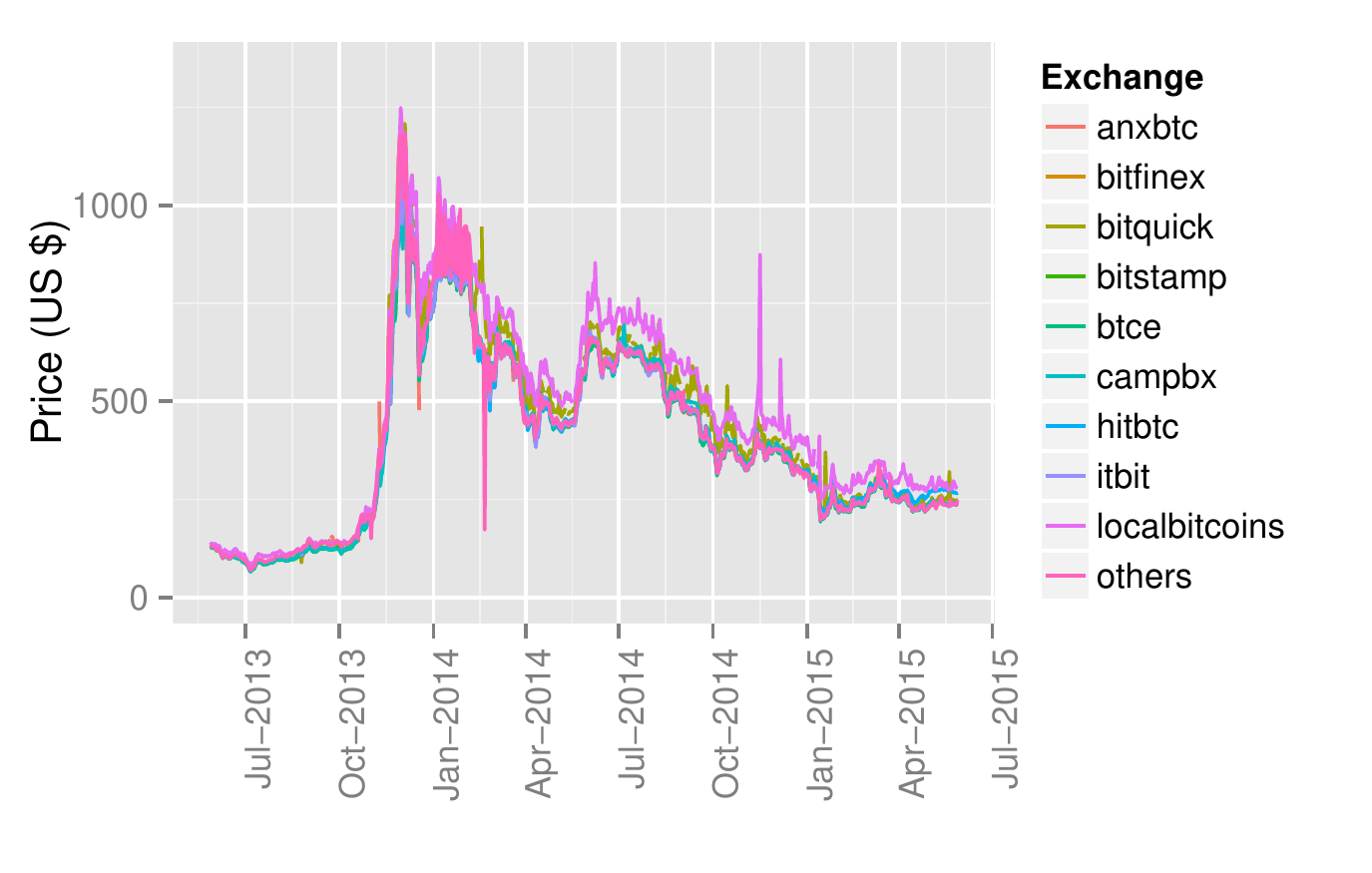}
\includegraphics[width=0.45\textwidth]{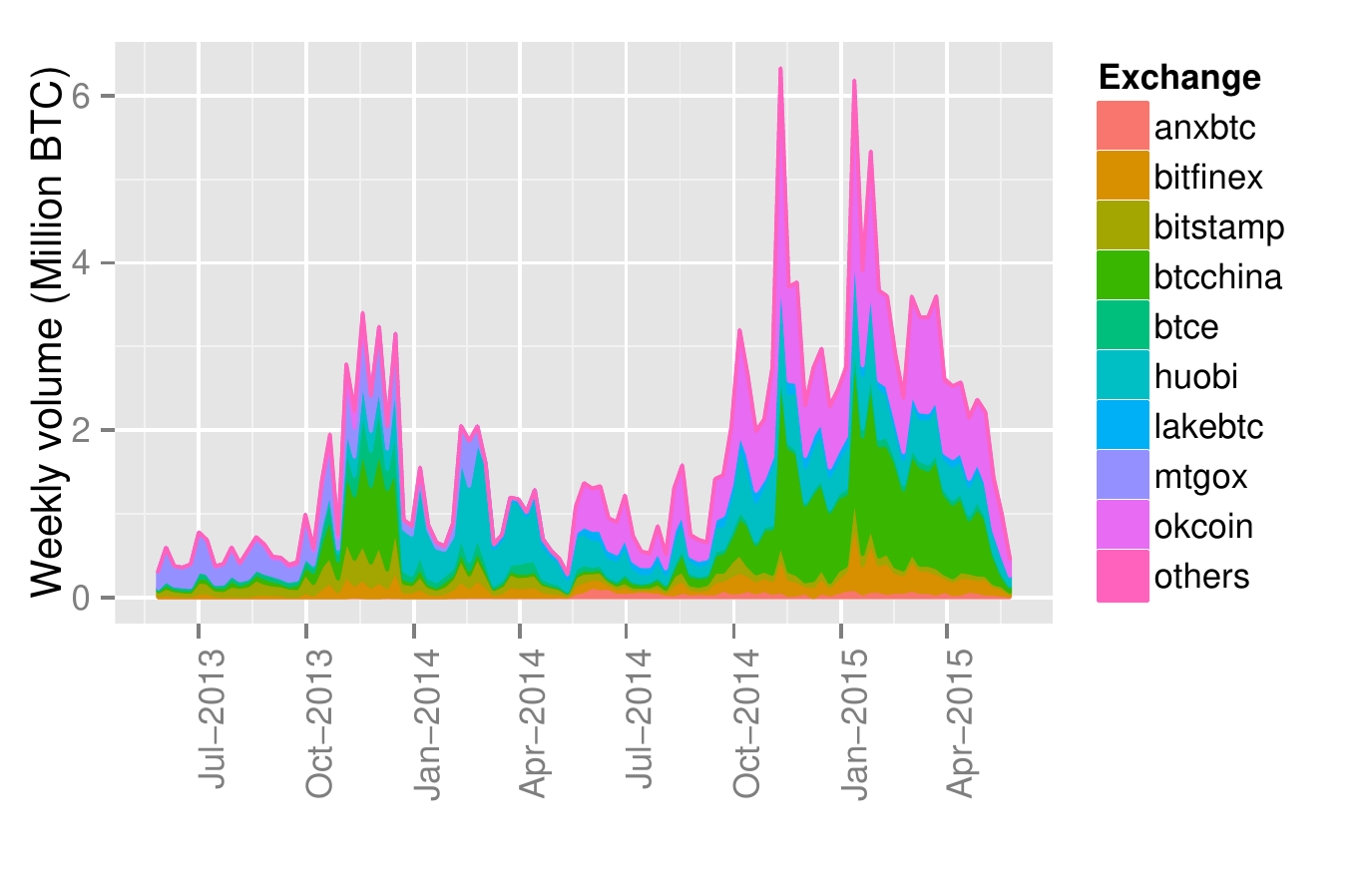}
\includegraphics[width=0.45\textwidth]{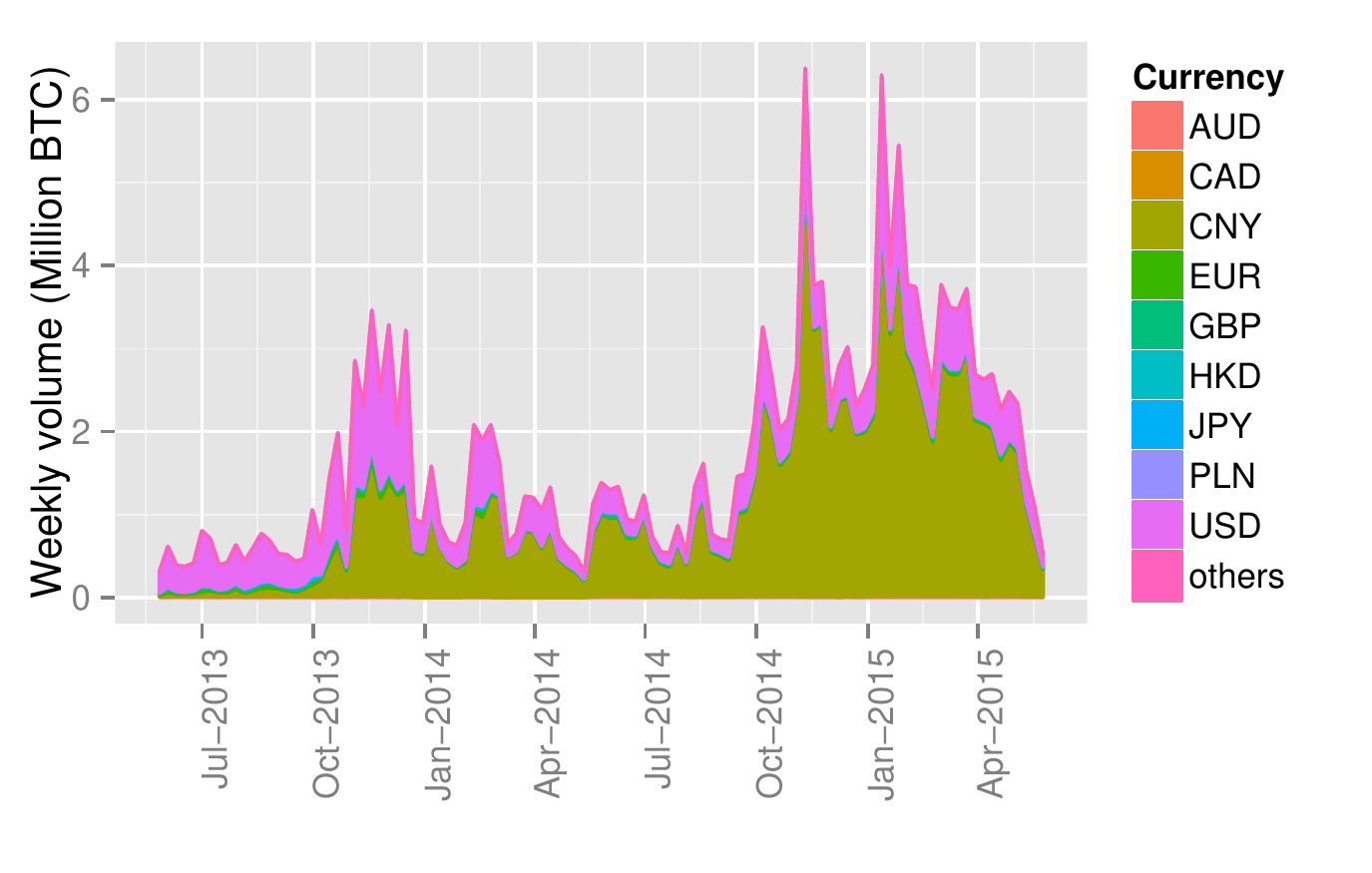}
	\caption{Top: Price fluctuation of Bitcoin over time. Bottom: Traded volumes by exchange (left) and by currency (right).}
	\label{fig:volumebyexchangecurrency}
	\end{center}
\end{figure}

Bitcoin is by no means the only crypto-currency. Coinmarketcap\footnote{\url{http://coinmarketcap.com/all/views/all/}, accessed 30/06/2015.} lists 590 currencies, with a total market capitalisation of \$4.5 billion. As Bitcoin accounts for more than 80\% of this amount, we will focus on it to exhibit trends in crypto-currency activity. Figure \ref{fig:volumebyexchangecurrency} shows the evolution of price, as well as traded volumes over a 2-year period. It is interesting to note that while trading in Bitcoin was predominantly in US dollars, it has now moved to being predominantly in Chinese Yuan. This highlights Bitcoin's nature as both a highly speculative investment and as a tool for evading currency controls\footnote{\url{http://www.ft.com/fastft/289502/bitcoin-still-gaining-currency-china}}.  

\begin{figure}[ht!]
	\begin{center}
\includegraphics[width=0.45\textwidth]{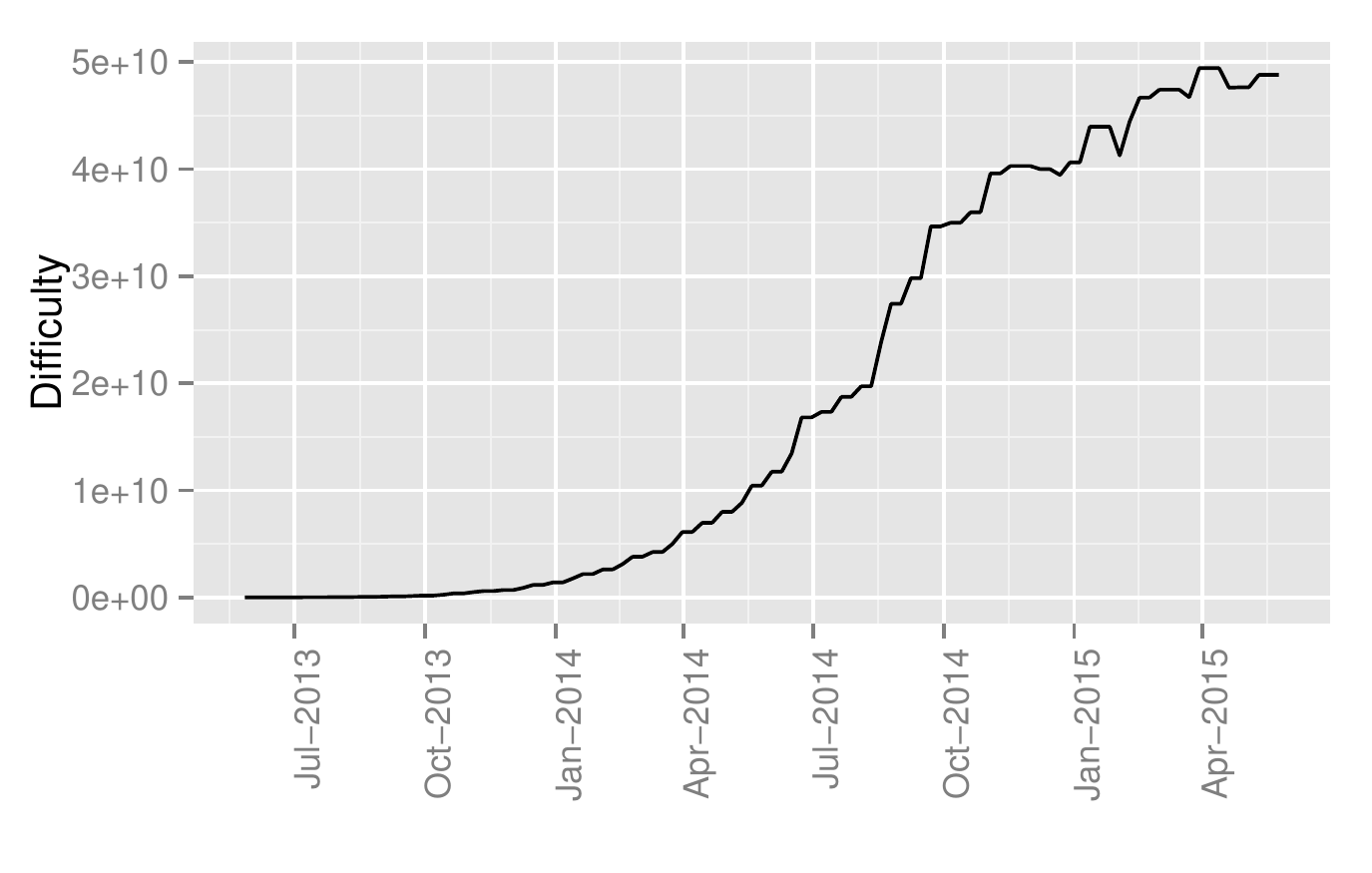}
\includegraphics[width=0.45\textwidth]{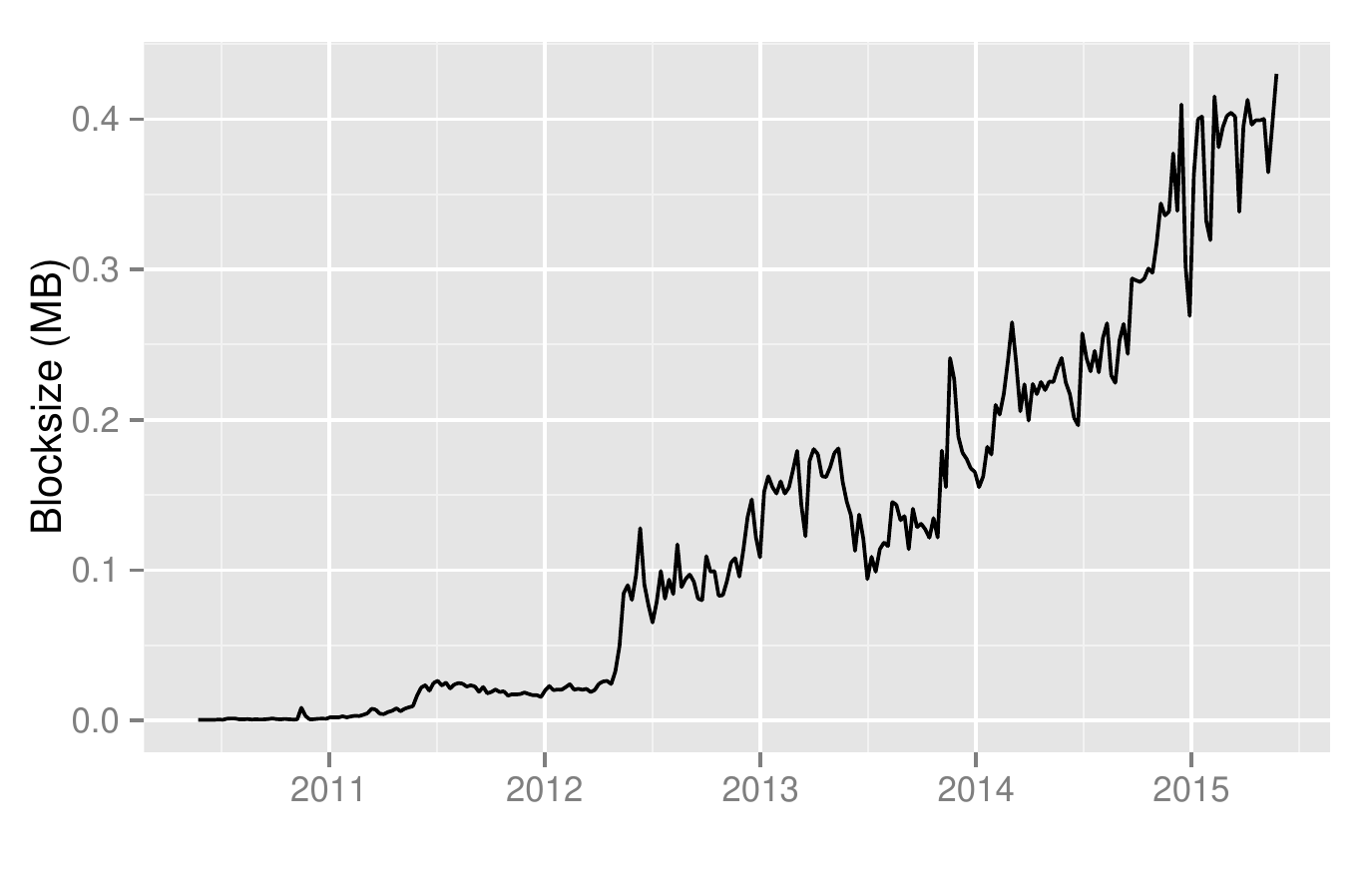}
	\caption{Difficulty}
	\label{fig:difficultyblocksize}
	\end{center}
\end{figure}

The Bitcoin network relies on `miners', or members that contribute computational power to solve a complex cryptographic problem and verify the transactions that have occurred over a short period of time (10 minutes). These transactions are then published as a block, and the miner who had first published the proof receives a reward (currently 25 bitcoins). The maximum block size is 1 MB, which corresponds to approximately 7 transactions per second. In order to ensure that blocks are published approximately every 10 minutes, the network automatically adjusts the difficulty of the cryptographic problem to be solved.    

Bitcoin mining requires specialised equipment, as well as substantial electricity costs, and miners thus have to balance their technology and energy investment so that their activities are profitable. As the price of Bitcoin increased, miners invested in more hardware, increasing their computational capability. However, the Bitcoin network then increased the difficulty of the cryptographic problem, in order to keep blocks published in regular intervals. Figure \ref{fig:difficultyblocksize} shows the evolution in both the difficulty of the cryptographic problem over time, as well as the block size. We note the exponential increase in the difficulty for a sustained period of time. As Bitcoin prices had been steadily declining in the latter part of this period, it is likely that mining became less profitable, which explains the plateau in difficulty. 

With regards to the increase in blocksize, this corresponds to an increase in Bitcoin transactions over time. A blocksize of 0.4 MB corresponds to approximately 3 Bitcoin transactions per second. 
A summary of other Bitcoin related trends is also provided in reports such as by \cite{bohme2015bitcoin}.

\subsection{Crypto-currency real world usage}
The projected future use of crypto-currencies like Bitcoin is discussed at length by \cite{brito2014bitcoin}, with regard to securities, options, swaptions, forwards, bonds that may be developed going forward based on virtual currencies such as Bitcoin. The European Central Bank, in its second report \footnote{\url{www.ecb.europa.eu/pub/pdf/other/virtualcurrencyschemesen.pdf}}, presents both an overview of the actors, the different modes of operation and the different business models that originate from virtual currencies schemes. Measures of current usage for Bitcoin shows between 60,000 and 70,000 transactions daily, for a total transacted volume of between \euro 15 and \euro 30 million, numbers which are somewhat insignificant compared to activity with existing payment solutions\footnote{Existing payment solutions include Visa, MasterCard, Paypal etc, and the ECB puts current daily non-cash payment transactions at 274 million.}. However, the ECB report highlights speed, cost and facilitation of cross-border payments as a major advantages of virtual currencies. 

The European Securities and Markets Authority (ESMA) has published a call for evidence on virtual currency investment products, as well as blockchain investment applications not involving virtual currencies\footnote{\url{http://www.esma.europa.eu/system/files/2015-532_call_for_evidence_on_virtual_currency_investment.pdf}}. This interest of ESMA is much more narrow than that of other stakeholders, in that it does not seek to express a view of the desirability of using virtual currency in a payment system. Instead, it focuses on collective investment schemes (CIS) and virtual currency derivatives. In its preliminary work, ESMA has obtained data from 6 of 13 virtual currency CIS, which had approximately \euro 246 million, with the largest accounting for almost half of this figure. Besides these schemes, ESMA also identified regulated European companies offering contracts for difference (CFDs) in Bitcoin and Litecoin, as well as binary options on either.

\subsection{Crypto-currency as a means of facilitating crime}
In its infancy, Bitcoin was associated with criminal activity through the online marketplace `Silk road', which operated on the Dark Web. Analysing 8 months of data from this marketplace, \cite{christin2013traveling} found that the majority of the 24,400 items sold on the market place were controlled substances and narcotics, with 112 sellers active throughout this interval. The total revenue from public listings in this time was approximately \$10 million. Silk road was shut down by the  FBI in 2013, while also seizing \$28.5 million in Bitcoin and arresting the marketplace's operator\footnote{\url{http://www.forbes.com/sites/andygreenberg/2013/10/25/fbi-says-its-seized-20-million-in-bitcoins-from-ross-ulbricht-alleged-owner-of-silk-road/}}. 

\cite{moser2013inquiry} provided the first thorough study of the potential for Bitcoin to be used as a money laundering tool. In particular, they investigated companies which provided anonymising services for a fee, by `mixing' Bitcoin inputs from several participants, and generating new Bitcoin addresses to hold the outputs. They determined that some services were indeed effective for this purpose and concluded that because of this, it is unlikely  that a Know-Your-Customer principle can be enforced in the Bitcoin system.  

In terms of real-world use in this context, an assessment of the National Crime Agency in the UK found that the majority of transactions for illicit purposes where actually of low value, and there was little to suggest that digital currencies have been widely used in the context of money laundering. Although anonymity was identified as a potential facilitator of criminality, in reality to use many of the available digital currency services, users would have to register an (eponymous) account. 

\subsection{Other distributed ledger technologies}
\label{sec:blockchain}
While HM Treasury and the Euro Banking Association (EBA) have been ambivalent towards Bitcoin in their recent reports, they have both recognised the potential of cryptotechnologies for other use cases. In particular, they have identified the distributed ledger at the core of the Bitcoin protocol, which achieves governance by consensus. While few concrete examples exist at present, \cite{swan2015blockchain} cites several examples of transnational groups which could use such a governance structure, such as the Internet Standards group ICANN and DNS, thus avoiding the influence (political and otherwise) of certain groups that would occur when registering in particular jurisdictions. A more ambitious example is that of smart property, where potentially every asset could be encoded onto this ledger with a unique identifier, and thus all asset transactions could be confirmed and tracked via the blockchain.

As noted in \cite{barber2012bitter}, the notion of scripting offered by crypto-currencies like Bitcoin is a highly useful and very innovative feature. It allows users to embed scripts in their Bitcoin transactions, this key feature is only just being recognised as a utility in its own right. It has been realized that at least in theory, as noted in \cite{barber2012bitter} that this can lead to ``... rich transactional semantics and contracts through scripts, such as deposits, escrow and dispute mediation, assurance contracts, including the use of external states, and so on.''

The Bitcoin use-case is one where the blockchain used is permissionless. `Permission' refers to the verifiers on the network, and in the case of Bitcoin, miners do not have to be authorised by a central authority before performing their mining activities. This is not the only model for a blockchain, however, and indeed the actors on the network who verify transactions can be subject to authorisation, as well as legal accountability. The applications outlined in this section span both modes of blockchain operation.

\begin{figure}[ht!]
	\begin{center}
\includegraphics[width=0.7\textwidth]{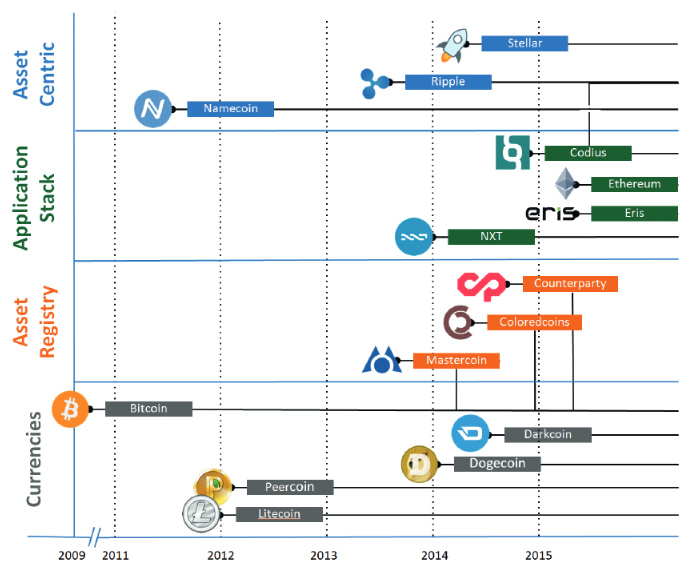}
	\caption{4 categories of cryptotechnologies. Reproduced from the EBA Cryptotechnologies report.}
	\label{fig:crypto}
	\end{center}
\end{figure}

In its report, the EBA \footnote{Available at \url{https://www.abe-eba.eu/downloads/knowledge-and-research/EBA_20150511_EBA_Cryptotechnologies_a_major_IT_innovation_v1.0.pdf}, accessed 29/05/2015} presents an analysis of cryptotechnologies in four application areas, presented also in Figure \ref{fig:crypto}:

\begin{itemize}
\item \textbf{Currencies} such as Bitcoin, Litecoin etc.
\item \textbf{Asset registries}: Similar to the smart property example mentioned earlier, ownership details would be recorded in the blockchain, and while physical assets could always be lost or stolen, the holder of an asset would not be able to claim ownership until it has been transferred via a blockchain transaction. However, because of the potentially large number of assets and associated details that could be recorded on the blockchain, this could create a large amount of traffic on the network. Bitcoin's 1MB block size caps the number of transactions at an average of 7 per second, and it is clear that a much higher number would be needed for the purpose of asset registration in certain areas (e.g. financial). A good example of a use-case is that of Everledger\footnote{\url{http://www.everledger.io/}}, a ledger for the certification and transaction history of diamonds. A laboratory first takes measurements of cut, clarity, size and other information and this is all stored on the blockchain.          
\item \textbf{Application stacks}: This application area aims to provide a platform for the execution of `complete applications on top of decentralised networks'. Examples include the smart contracts proposed by by Eris Industries \footnote{\url{https://erisindustries.com/}}, which can automatically verify the interactions between the parties to the contract. With such contracts, there is the possibility of creating derivatives that settle automatically and reduce counterparty risk, such as the blockchain derivatives developed by Hedgy\footnote{\url{ http://hedgy.co/}}. There are several caveats to this application area also, however, as smart contracts will always be limited to the ability of the data to describe these interactions.
\item \textbf{Asset-centric technologies}: These focus on digital representation of real assets on a shared, but not public, ledger. 
\end{itemize}

\section{Value generation in crypto-currency}
\label{sec:value}
At first glance, it may be difficult to comprehend why crypto-currency, as a purely artificial digital construct produced as a result of solving a computational problem, with no backing from a central authority, should have any value in the real economy. In this section we will refer to a number of economic principles followed by associated monetary theories, in order to determine any elements which could explain the value of this digital resource. We note that we do not advocate one particular school of economic thought over another, but will rather discuss issues that may arise under a range of these different prospective analytical frameworks, if crypto-currencies were to interact more widely with the real economy.

\subsection{Crypto-currencies as scarce economic goods and the potential of a `Deflationary Spiral'}
\label{sec:scarcity}
\cite{graf2015commodity} suggests that bitcoin `meets key characteristics of a good, as defined in relation to action and choice'. It is in fact a scarce digital good, produced through a predetermined issuance process, and guaranteed not to exceed a certain quantity, as its protocol has a hardcoded upper limit of 21 million coins, a kind of asymptotic upper bound. While one is accustomed to think about goods and scarcity in a material sense, this of course does not have to be the case. 
 
Consequently, it is then worth considering what will be the final means of value generation when the money supply for instance in Bitcoin is complete, either by means of exhausting the computational effort one is willing to expend in mining more coins or the actual total number of Bitcoins is produced. Unlike physical metal commodities, which are in unknown total supply, we argue that the knowledge of the total amount available will change the perceived value of the currency. Though physical metals may be scarce, the lack of knowledge of their total supply leads an ever more involved and expensive search for more, maintaining or increasing the worth of those currently in circulation, this will not be the case with Bitcoin. At which point the argument of value maintenance for such a crypto-currency must change to a different perspective. 

Some economists, such as Paul Krugman\footnote{\url{http://krugman.blogs.nytimes.com/2011/09/07/golden-cyberfetters/}} observed the following possibility of deflationary pressure in crypto-currency networks. Bitcoin's capped total money supply could be viewed as a variation on Milton Friedman’s ``k-percent rule'' \citep{friedman1960program}. This theory states that an optimal way to control inflation over the long term is for the central bank to grow the money supply by a fixed amount of k\% each year, irrespective of the cyclical state of the economy. In particular, one should set the growth variable of k\% at a rate equal to the growth of real GDP each year. This connection between Milton Freidman's Nobel prize winning theory and Bitcoin practice was highlighted recently in \cite{bohme2015bitcoin} who argue that one can consider Bitcoin as a type of ``... proposal to fix the annual growth rate of the money supply to a fixed rate of growth.'' At the end of the mining process, when the total Bitcoin money supply is created, this would be equivalent to a $k=0$ or perhaps a negative $k$ if a large loss of money supply occured due to theft, electronic storage corruption or damage to physical storage of a non-trivial portion of the total money supply.

Hence, one needs to consider what is applicable monetary policy to deal with the situation that the size of an economy grows at a different rate to the quantity of money in that economy, in this case Bitcoins. \cite{bohme2015bitcoin} reiterate the views of Paul Krugman that ``... the fixed slow growth rate of Bitcoin creates the possibility of deflation if Bitcoin was to be used widely...''. They also note that there have been other crypto-currency extensions of Bitcoin proposed to overcome such potential problems, see discussion by for instance \cite{king2013primecoin} which introduces Primecoin with infinite money supply or the introduction of Peercoin which keeps k\% around 1-2.

\cite{barber2012bitter} also discuss such issues, talking about a deflationary spiral that may arise from the capped money supply. We first briefly recall what a deflationary spiral is before discussing this in the context of Bitcoin.

A deflationary spiral refers to an economic development where rampant deflation can eventually lead to the collapse of the currency. In general deflation can be considered as a decline in the general price level. It can occur when the price of goods and services, as measured relative to a specific measure, begin to decline. This may not be due to the fact that the value of the goods and services themselves reduced, instead it can simply occur due to the fact that the value of the currency itself increased. So one can consider the spiral of deflation as arising in the situation that the value of a currency, relative to the goods in an economy, increases continually as a result of hoarding. In response, as the value of the currency relative to the goods in the economy increases, people are given an incentive to hoard the currency. This incentive arises from the fact that by retaining the currency, they aim to be able to purchase more goods for less money in the future, this becomes a vicious cycle as the lack of available currency in the economy causes prices of goods to decrease and this results in yet further hoarding.

Such an effect is a real condition that affects the fiat backed fractional reserve banking system. There are two schools of thought as to whether such a deflationary spiral may occur for Bitcoin. One view is that it is not likely to occur in the case of Bitcoin, since it is argued that users in the real economy may not foresee a fixed cost (unit amount) that they must pay with Bitcoin. Therefore, if the value of the Bitcoins that they own increases, then one may expect that any future cost will take a proportionally smaller amount of Bitcoins. A consequence of this view is that there would however by no real fixed incentive to hold Bitcoin other than pure speculation.
In addition, if the real economy that allows Bitcoin grows, then one would also expect the per-unit value of Bitcoin in such a perspective to proportionally increase. This view effectively perceives Bitcoin not as a debt but as an asset, and as such under such a perspective one would expect that Bitcoins would only deflate in value when the Bitcoin economy is growing.

In \cite{barber2012bitter} they take this perspective and they postulate on a setting in which Bitcoin usage has matured in the real economy, considering for instance a stable 1\% of US GDP transactions in Bitcoins and 99\% in USD. They then argue that in such a setting one may expect that the purchasing power of Bitcoin would still increase over time. The reason is that each coin will increasingly capture a correspondingly constant fraction of the country’s growing wealth. They acknowledge that such a deflationary spiral may occur for bitcoins and discuss potential for hoarding of such crypto-currency. They argue that their appreciation potential will result in a user tendency to accumulate Bitcoins rather than spend them in the real economy. The consequence of this is that the incentives offered to groups that verify and validate Bitcoin transactions on the blockchain will reduce as there will be less Bitcoins in circulation, hence transaction volumes naturally reduce resulting in a less profitable operating environment for verification of transactions. They aptly term this condition ``bit rot''.

The alternative economic perspective on how deflationary spirals may manifest is given by the argument that they occur when there is an incentive to hoard because of declining prices. The decline in prices will result in less available currency in the market place, which further perpetuates a decline in prices, and the deflationary cycle emerges. The website \url{https://en.bitcoin.it/wiki/Deflationary_spiral} discusses mechanisms under which a non-traditional deflationary spiral may arise in the Bitcoin network. It argues that once Bitcoin value stabilizes there will always be the knowledge that the number of Bitcoins in the market is limited. Consequently, if the total value of all Bitcoin transactions completed increases in "real" terms, then there will continue to be price deflation. From this view, there can be an expectation of future deflation which will result in a discrepancy in perceived values of Bitcoins depending on ones investment horizon. In the short term under this scenario, there would be an apparent over-pricing of Bitcoin, which may encourage alternative competition.

\subsection{The metalist view}
\label{sec:metal}
A range of authors have alluded to the metalist perspective on understanding the value generation mechanism for the Bitcoin crypto-currency, see discussions in \citet{maurer2013perhaps,ingham2004emergence,blanchette2011material}. For instance, \cite{maurer2013perhaps} discuss Bitcoin and the embracement of its users in a form of monetary pragmatism, and state ``... Bitcoin enthusiasts make the move from discourse to practice in their insistence that privacy, labor, and value are ‘‘built into’’ the currency’s networked protocols. This semiotics replays debates not just about privacy and individual liberty, but about the nature of money, as a
material commodity or chain of credits.''. They argue that Bitcoin embodies a form of ``practical materialism'' which is manifest in the form of a modern day digital metallism, an extension of the ideas of \cite{ingham2004emergence} and his perspectives on ``practical metallism''. 

Both \cite{blanchette2011material} and \cite{maurer2013perhaps} argue for a form of metalist monetary perspective on Bitcoin. The latter stating ``... Despite the supposed immateriality of digital bits of information, matter itself is very much at issue with Bitcoin, both in how it is conceptualized and in how individual Bitcoins are ‘‘mined.’’...''

Under the premise of a ``metalist's'' view of the value derivation of money, many would argue that value of crypto-currencies may at present be derived from physical commodities consumed in the mining process utilised to obtain this increasingly scarce resource. For instance several studies have argued that the price of crypto-currency Bitcoin is related to the cost of maintenance, storage and electricity consumption required for the large server farms ``virtual mines'' utilised to create the bitcoin currency, see discussions in \cite{o2014bitcoin}. In \cite{JP2011bits} they argue that the material value of Bitcoin is not limited to the privacy feature offered by the crypto-currency, they argue that it finds another feature that provides its value, the process of producing new Bitcoins known as mining which ``mimic[s] the extraction of minerals [...]. As the most readily available resources are exhausted, the supply dwindles.'' 

If one then continued the perspective of a metalist monetary theory for crypto-currency such as Bitcoin, then one could argue based on ideals expressed in \cite{ingham2000babylonian}, where they consider money to be the consequence of rational agents that prefer to work with money that is the most tradeable commodity in the current real economy. Under this perspective there is some notion that virtual and crypto-currencies especially could maintain value after the mining process. For instance, if rational agents in the economy began to prefer or value them more than other fiat backed e-money substitutes. This could happen in a number of ways, for instance rational agents may prefer the privacy features such virtual and crypto-currencies may offer in the digital economy more than other fiat based e-money competitors. Another possibility may be that the block-chain technologies that act as ledger, for instance in Bitcoin, may find wide-spread uptake as a means of virtual contract construction between different economies, or as a third perspective, if virtual and crypto-currencies found a wider market base in third world countries by moving beyond internet based services to mobile services, this may also maintain their value in the real economy.

\subsection{The chartal view}
Next we discuss some alternative monetary theory perspectives on cryto-currencies such as Bitcoin. In particular, we consider the case of Bitcoin when the mining process is completed and all the money supply has been created. We then consider the chatalist perspective of where Bitcoin may derive its value, this is an alternative perspective to that of the metalist views expressed above that has not been discussed previously in the context of Bitcoin. Therefore, we find it interesting to open up this avenue of thought to more debate.

An alternative view to the metalist perspective can also be considered, where the value of Bitcoins may continue to be maintained. This alternative view would be based on a transition from the metalist perspective, post mining completion, to a chartalist's view. This view posits that money should not be studied in isolation from the powers of the state, i.e. the country that ``created'' and ``controls'' the money. In particular, under this perspective, money in its general sense is a unit of account created by a central (government) authority for the legal structuring of its social debt obligations.

Well before crypto-currencies were conceived of, for instance \citet{knapp1924state} argued that all monies are chartal, and this can include crypto-currencies, since all payments in the form of tax to the state or governing powers are measured in some unit of value. Furthermore, the state makes a decision ``that a piece of such and such a description shall be valid as so many units of value'', it is then irrelevant what this token or money manifests as since it is only a ``sign-bearing'' object that a state ``gives a use independent of its material''.

\subsection{How do 'outside monies' like virtual and crypto-currencies fit into the chartal and modern monetary theory perspectives ?}
In this section we delve in more detail into the importance of thinking about the role of such virtual and crypto-currencies in aspects of monetary theory and monetary policies if they become more prevalent in the real economy. We contrast views formed based on fiat backed e-money with the how they may be affected in a real economy with both fiat and virtual or crytpo-currencies. In general we will tend to raise more questions, than we proffer solutions. Though this is useful to open dialogue and ways of thinking about the challenges that may lie ahead. 

In particular we first recall that monetary theory is developed with the aim of understanding the most suitable approaches to monetary policy and how it should be conducted within an economy. It is suggested by such theories that a variety of different monetary polices may be employed to benefit countries, depending on their economy and resources. For most monetary theories, the core ideals relate to factors such as the size of the money supply, price levels and benchmark interest rates and how they all affect the economy through inflation, taxation, wage growth and unemployment levels. It is then the realms of economists and central bankers to execute the outcomes of such theories in practice. As stated, we would like to initiate some exploration of how virtual and crypto-currencies, when mixed with fiat currency in the real economy, may alter traditional outcomes on policy decisions compared to fiat backed money supplies.

There are many forms of monetary theories that have been developed by economists, indeed we have seen brief discussions on metalist and chartal views already above. These include ideas of Fiat Debt-Free Money Reformers, Modern Monetary Theorists, Modern Monetary Realists, Post Keynesian Reformers, Islamic Banking Advocates, Social Credit Reformers, Land Reformers, Hard Money Reformers and Competing Currency Reformers. Recent, some would say unorthodox versions of such theories \cite{tcherneva2006chartalism}, include variants such as Modern Monetary Theory (MMT) \cite{wray1998understanding} and Modern Monetary Realism (MMR) which were developments from early forms of Chartalism \cite{wray1998money} and prior ideas from \cite{knapp1924state,forstater1999functional} and functional finance theories of \cite{lerner1943functional}. Such theories also are termed neochartalist approaches and ``tax-driven'' money, see discussion in \cite{wray2000neo}. All these theories revolve around the procedures and consequences of utilization of government issued units of money often called fiat money, in the sense of the definition offered earlier.

A key premise of theories like MMT and the consequences of monetary policy that flows from these theories is the notion that governments have some level of control over the money supply and elasticity of money. So we wonder, what happens to such controls when other forms of currency, created outside of any sovereign state, starts to interact in a given economy. Does this reduce the power of the state to enact policies based on the assumption of ultimate control of money supply, or does it act as a friction or damping factor on the utility of resulting policy levers when enacting policies assuming ultimate money supply controls are still relevant. 

One can view money, in its general sense, as a unit of account created by a central (government) authority for the legal structuring of its social debt obligations. For instance, this may manifest between a population and a governing central figure in the form of taxation liabilities. In this setting it is conceived by chartalists and many modern monetary theories that money then arises from the state as a form of tax credit that can nullify these taxation debts. This is in firm contradiction to other orthodox theories that followed from commodity based currency views such as gold standards which view money more as naturally arising as a medium of exchange from the attempts of enterprising individuals to minimize transactions costs in barter economies.

No matter which view one prefers, it is interesting to question what implications may arise from interactions in such economies of non-government controlled currencies which are non-fiat such as virtual currencies and crypto-currencies acting as truly ``outside'' monies. Before embarking on developing such questions for future consideration we summarise a few key ideas from chartalist, MMT and MMR thinking, based on the account provided in \cite{tcherneva2006chartalism} where it is observed that in general the following principles are considered by these theories. With each concept, we briefly pose questions relating to their applicability in the setting of an economy which admist both fiat currency as well as virtual and crypto-currencies.
\begin{itemize}
\item{Dismissal of the view that money emerges naturally as a medium of exchange that enables the minimization of transaction costs among utility maximizing rational agents in the real economy, due to their view that such notions lack historical support.
}
\begin{itemize}
\item{\textsl{Is this view now valid for crypto-currencies. Some would argue one of the key reasons crypto-currencies are being adopted in the real economy at present is due to the very fact that they are providing a reduction in transaction costs for some agents in comparison to other fiat backed e-money payment services such as Paypal, see discussions in \cite{brito2014bitcoin}. Perhaps, therefore there will be some historical precedent for questioning this perspective further in the case of virtual and crypto-currencies.}
}
\end{itemize}
\item{One should study money in the context of institutions and culture with special consideration given to political and social considerations.
}
\begin{itemize}
\item{\textsl{Certainly, the role of virtual and crypto-currencies may fit into this perspective, in the sense that the context of their uptake in the real economy has historically certainly been a function of institutional influence from governments in the form of regulations and central bank policies. The role of virtual and crypto-currencies has also been influenced by cultural and social considerations. To see this one may consider for instance the rapid uptake of some virtual and crypto-currencies in the U.S. and more recently in China, where in some cases they are used as alternative means for transmission of assets with enhanced anonymity from central government oversights.}
}
\end{itemize}
\item{Money is by its nature a credit-debt social construct. Furthermore, chartalists argue that social debt relationships may be ordered with the top of the hierarchy being the liability of the central authority which they deem the most reliable. Neochartalists also argue that modern currencies are contained in a context of certain governing central or state controls: the ability to levy taxes on the population and economy; and the ability to decide what is acceptable for payment of tax liabilities. In this context tax should be understood in a broader context of modern income tax, estate and commercial tax as well as any non-reciprocol obligation to the state such as fines and fees.
}
\begin{itemize}
\item{\textsl{We will address this point in Section \ref{Acceptance} }
}
\end{itemize}
\item{Money functions as an abstract unit of account which is then used as a means of payment and debt settlement. Unlike orthodox monetary theories, charatalists distinguish between money-of-account and money in the real economy, perhaps summarised by \cite{keynes1930treatise} who argued that ``money-of-account is the description or title and the money is the thing which answers the description.'' With this view, chartalists see money's function in the real economy as a medium of exchange is incidental to and contingent on its primary function as a unit of account and a means of payment of liability. Neochartalism generally views taxation not as a form of financing government spending but instead as a mechanism to create demand for the currency.
}
\begin{itemize}
\item{\textsl{We will address this point in Section \ref{Acceptance} }
}
\end{itemize}
\item{Neochartalists believe that given the view that modern states or countries or unions have the monopoly power over the issue of their currency, i.e. sovereign currency control with no fixed exchange rates, dollarization, monetary unions or currency boards, they they will not face operational financial constraints, though they could face political constraints. Furthermore, they consider that such states should consider borrowing as an \textit{ex ante} interest rate maintenance operation, arguing that instead the taxation system is established as a means to creating demand for currency rather than financing of government spending.

Their perspective is such that, no entity with the power to create and destroy money as they require will need anyone else to assist in the ability to `fund' spending. However, even though deficits for the economy are not financially constrained in the typical sense, they are still subject to potential pressures from inflation rates and exchange rates, as well other considerations such as access to available resources, capacity utilization, labour availablility, and external balance. 
}
\begin{itemize}
\item{\textsl{Firstly, we discuss the issue of monopoly power over currency supply. To address this consideration, the question that may arise is whether or not the central bank or government can control the money supply and elasticity of such decentralised virtual or crypto-currencies perhaps through accumulation of stored reserves raised through taxation? This would ofcourse be assuming they were eventually allowed by governments as alternative forms of payment for tax liabilities along side traditional fiat currency. If this were the case, then one would need to be very careful in the money supply management, since as noted previously too greater hoarding of these currencies, which are of bounded total money supply, may result in a deflationary spiral.
}
}
\item{\textsl{
An alternative perspective, which avoids the need for reserving of virtual or crypto-currencies, in order to achieve control of the money supply may also be possible for some types of virtual and crypto-currencies. For instance, in the case of Bitcoin, instead or accumulating reserves, a government may alternatively take greater stakes in the network mining and transaction validation activities. A governments access to vast computing power, relative to most agents in the economy, puts them at a distinct advantage to gain sufficient computational power within such networks that any virtual or crypto-currency with consensus network type protocol embedded in its code may be able to have its core attributes modified by governments who earn sufficient voting rights. For instance, a government may gain sufficient control of the currencies network to alter core features of the code such as the finite money supply aspect, the mining rates and other key features related to the money supply. Perhaps it may be argued that in effect this is the crypto-currency equivalent of state central power over money supply.
}
}
\item{\textsl{Secondly, we consider the issue of whether virtual and crypto-currencies would result in a form of operational financial constraint for states and governments. In the case of decentralised virtual or crypto-currencies, the operations required to gain some form of control or assert some form of management of the money supplies in the real economy may not in general be free from operational financial constraints. For instance the actions mentioned above such as reserving of virtual or crypto currencies, or more active control/'voting power' with in the virtual network, through enhanced mining or transaction processing activities will be potentially expensive for the state to maintain and can be considered as a operational financial constraint on the actions they may wish to enact in their monetary and fiscal policies.
}
}
\end{itemize}
\item{Neochartalists also consider that when a state has a monopoly over the currency, it also has the power to set prices, including interest rates and how currency will be exchanged for other goods and services.}
\begin{itemize}
\item{\textsl{So if one assumes that the state only has partial power over some aspect of a virtual or crypto-currency through such means as discussed in the previous bullet points above, then an interesting question to raise is what implications does this have for the perspectives held by Neochartalists views on the ability of a state to set prices, interest rates and exchange rates? These views are based on the premise that the state has monopoly power over the currency, and ofcourse they will still maintain this over their fiat denominations. So the point of consideration is more whether an increased growth and uptake in the economy of virtual and crypto-currencies, for which the state does not have monopoly control over the money supply attributes, will create a friction in their ability to set prices, interest rates and exchange rates ?  
 }
}
\end{itemize}
\end{itemize}

\subsection{Acceptance and Legal Tender} 
\label{Acceptance}
Many have argued against various aspects of MMT and related theories from a chartalist root. One of the key aspects they point to relates to the notions of legal tender. For instance, \cite{schumpeter1954history} and \cite{davidson1972money} emphasized legal tender laws as critical, where the state or government would issue a currency in terms of a unit of account and then pass laws to require adoption of that currency in designated public and private payments. This is a jurisprudence perspective of how currency can become valuable in a real economy, however chartalists like \cite{knapp1924state} took an alternative view that such laws would not suffice and that the state or government effectively establishes the money of account when it determines what will be ``...accepted at public pay offices...'', rather than through legislation. 

Hence, we see that an important point to note that is directly of consequence to understanding a chartalist's view of virtual currency and crypto-currency is to observe that the chartal nature of money and its acceptance in the real economy lies not in its acceptance in the form of a legal tender status but instead on its place in the heirarchy order of social debt relationships. This derives instead from the states power to delegate taxes and dictate how and in what form of money such accounts will be paid. 

Therefore, under a chartalists view on monetary theory it is not a question of whether fiat currency is in direct competition with virtual or crypto-currencies, but instead whether there will be sufficient demand from the public that will enforce the will of the public to push the state to accept such currency forms as means of payment of liabilities owed to the government. Should this occur, there will then be an interesting circumstance arising where one unit of account is established in a fiat currency which is under the control of the government, however a second unit of account is from a decentralised money supply mechanism in the form of crypto-currency. We point out that has potential to change dynamics in the supply and demand of fiat currency and should be considered further.

\subsection{Competition between virtual/crypto-currencies and fiat backed currencies}
\label{Competition}
Another interesting point to make that arises naturally from a chartalist view and relates to virtual and crypto-currencies in regards to the concern some have raised about such monies competing and perhaps becoming a dominant unit of barter in an economy is that agents can never simply refuse to take a sovereign's money. That is, fiat currency is the key money to make payment for taxation liabilities, so long as there is always taxation present in the economy, which in some form relies upon the fiat currency more than the virtual or crypto-currency. In this case, the fiat currency will always remain at the top of the hierarchy of social order in terms of debt relationships, see further discussion on this general view in \cite{tcherneva2006chartalism}. The only issue arising in such cases is again the fact that when virtual and crypto-currencies are allowed into the economy to pay tax they diminish the power of the state to posses and maintain unconditional control of the currency, that they would maintain if they only allowed for receipt of tax credit their own unit of money or fiat currency. 

Consequently another issue arises here that potentially complicates the above considerations, this is the one pointed out by Innes \cite{innes2004credit} where it is argued that it is not only the requirement to pay taxes in any particular state mandated monies, but also the difficulty in obtaining these monies that provide the monies worth. To understand where this may pose a challenge to fiat currencies, one needs to consider the situation in which fiat money and virtual and crypto-currencies are allowed in the economy (perhaps not as legal tender) but to settle tax debt in government offices. In this case, if it is perceived by the public that certain attributes, for instance privacy features of virtual currencies or crypto-currencies are more valued that those of fiat denominated e-money, then it may be conceivable that these would have preference in the economy. Now add to this the scarcity of such bitcoin monies in terms of the hard limit on their physical creation, unlike government money which is only really limited by inflationary pressures in the given economy and one has an interesting question to postulate relating to which form of currency and in what conditions would maintain the top heirarchy in terms of social debt settlement unit.

\subsection{Not high powered money and yet somehow explicitly liability free ?}
\label{Highpowered}
Consider the context of a modern economy with a fractional banking system in place. In such an economy, a bank recognizes that it is safe to issue deposits to an amount that is some multiplier of its actual physical reserves since it may be reasonable to expect that only a small fraction of depositors will try to "cash-out" deposits, redeeming them for reserves. Then, under the setting in which a reasonably stable deposit multiplier is established as a function of the ratio of reserves held against deposits, the supply of deposits will then be determined by the quantity of loans demanded and the quantity of reserves supplied. One can then consider the role of governments in controlling this process, they are effectively able to exert some measure of control by deciding what should form the basis of reserves and also by establishing a legally required reserve ratio. At some stage this corresponded to the gold standard and now days has moved instead to government fiat money sometimes known as a form of High Powered Money. Since the government then has the ability to control the fiat money supply i.e. a seigniorage in the real economy, they then naturally obtain a level of control in the economy since banks will continue to have a demand for such currency in order to increase the value of their loan books, which is constrained by their ability to accumulate reserves and a reserve ratio condition on lending.

Hence, a modern economy revolves around a money supply that consists of bank deposits plus the portion of high powered money created by government that is not held by banks as reserves. Even though the banks may exert some level of control on the amount of fiat money help by the general public by adjusting interest rates on deposits to induce them to deposit or spend fiat money, however the government with its control of high powered money supplies to banks and it setting of reserve ratios, exerts exogenously a pressure on banks and ultimately the money supply.

Hence, another point worth questioning is the role of these exogenous currencies like virtual and crypto-currencies which are not created by central banks or private banks. Somehow they are liability free in some sense and yet they may not be considered in the Neo-Chartalists view as High Powered Monies, issued by central banks for spending in the private sector to fuel taxation generation and value creation in fiat currency. Unlike the view that although banks can also create money, their creation is a ``horizontal transaction'' since such created credit or money does not increase net financial assets since these assets are offset by liabilities. However, this is not the case with virtual and crypto-currencies, in addition, if they were allowed as monies to make payment for taxes and fines from a given government, then their legal power to discharge debt would increase their worth. This may cause a friction with the fiat denominated e-money system, since unlike fiat e-money which is issued or controlled by the government where it can issue its own currency at will subject to a public liability in the countries accounts appearing as a deficit in the countries accounts, it has no control over the issuance of the virtual or crypto-currencies except that which it may exert should it store significant reserves of such currencies in the central bank. This may therefore in principle, should virtual currencies become more mainstream act as a problem for the universality of the policy tool governments have utilised for years based on their universal monopoly of money creation that regulates inflation and unemployment.

In continuation of these above lines of questioning, one would wonder about the government or states ability to utilise money creation and taxation to control the rate of spending in the economy and therefore the ability to fulfill, as \cite{lerner1943functional} puts it, ``...to fill its two great responsibilities, the prevention of depression, and the maintenance of the value of money''. If virtual currency or crypto-currency were to be admitted as viable tender to pay tax to the government, then such currencies may diminish the standard monetary controls available to the government, since currency creation is no longer the sole mandate of the government, it would therefore require some form of symbiotic relationship with the fiat money supply and the virtual or crypto-currency supply to maintain the status quo. A fact that has not been lost on central banks over the years as early forms of e-money and non-fiate monies arose. 

One last point to make about the notion of liabilty in the case of virtual and crypto-currencies is perhaps that they are implicitly creating liabilities. This can be seen in the case that in the creation of such currencies through the mining process, the mines require utilisation of resources, loans/credit agreements with banks in creation of resources required to run and setup such mines, meaning the implicitly the creation of such currency, though explicitly it seems liability free, is actually implicitly not free of liability. 

\section{Views on crypto-currency from a regulation perspective}
\label{sec:regulation}
Given the importance of understanding the role of crypto-currencies in the monetary system highlighted above, we now turn to another core element that must be considered should such currencies be utilised increasingly in the real economy, the role of regulation. A detailed account of several aspects of regulation response to crypto-currencies can be found in \cite{peters2014opening}.

Even before the advent of crypto-currencies, there have been concerns about how centralised virtual currencies may limit a country's ability to control inflationary pressures. The Chinese Q-coin was adopted widely as a form of payment by online entrepreneurs, i.e. outside the online messaging environment which it was created for. The Chinese central bank, citing concerns about an increased money supply outside of its control, as well as a difficulty in imposing taxation, enacted limits in the issuance of these currencies \cite{lehdonvirta2014virtual}.   

A number of regulators around the world have been devoting an increasing amount of attention to virtual and crypto-currencies in recent years. \cite{mitchell2014bitcoin} outlines the responses of several regulators, from which one can observed that there are both varied interpretations of crypto-currency (e.g. as e-money, private money\footnote{Bitcoin has been recognized by the German Finance ministry as a unit of account, and is thus treated as a type of private money. \url{http://www.spiegel.de/international/business/germany-declares-bitcoins-to-be-a-unit-of-account-a-917525.html}}, as a commodity or private property, or as a private unit of account), which informs their treatment from a taxation perspective also. In most regulatory responses to virtual currencies in Europe, Bitcoin has not been found to fulfill the criteria/definitions of a currency. Sweden however has required virtual currency exchanges to register with the financial supervisor, while Germany and France have declared that certain Bitcoin related activities are subject to authorisation. There is no unified approach to regulation of such virtual currencies as payment services within the EU, and the European Central Bank (ECB) has not expressed any intention to amend the current legal framework to incorporate such considerations. We will discuss in a little more detail the recent responses of the ECB and the UK's HM Treasury, who have both conducted surveys about the use, benefits and risks of virtual currencies, as well as the New York Federal Reserve's recently released detailed regulatory framework. 

In November 2014, HM Treasury in the UK issued a call for information, attracting over 120 responses from diverse participants, including banks, payment service providers and digital currency developers. Results were published in March 2015\footnote{\url{https://www.gov.uk/government/uploads/system/uploads/attachment_data/file/414040/digital_currencies_response_to_call_for_information_final_changes.pdf}}.
Benefits of digital currencies include lower costs and faster, 24 hour processing availability, particularly for cross-border transactions. The risk side of these advantages are limited controls over transactions, theoretically allowing very large international transfers, with no capacity for the authorities to freeze or reverse payments, given the irreversibility of transactions in virtual currencies.

The ECB has been actively considering monetary policy implications resulting from the introduction centralised virtual currencies and decentralised crypto-currencies since at least 2012. In its first report\footnote{\url{https://www.ecb.europa.eu/pub/pdf/other/virtualcurrencyschemes201210en.pdf}}   it noted that both virtual currencies and crypto-currencies fall under the responsibility of central banks, due to the characteristics shared with payments systems, it highlighted the lack of supervision and concluded that they did not pose a risk to financial stability. In its more recent study\footnote{\url{www.ecb.europa.eu/pub/pdf/other/virtualcurrencyschemesen.pdf}},, it suggested thatd ue to its high price volatility and low acceptance rate, the Bitcoin could not be, yet at least, regarded as a full form of money from an economic perspective. The ECB revised its definition of virtual currency as `a digital representation of value, not issued by a central bank, credit institution or e-money institution, which, in some circumstances, can be used as an alternative to money'. 

Despite the slow uptake of virtual currencies, the ECB also has stated its intention to monitor possible threats to monetary policy and financial stability, in the case where virtual currencies gain mainstream acceptance. It suggests that this would be possible for a new generation of virtual currencies which address current technical weaknesses and are geared towards a more mainstream, less technologically minded audience.  

With regards to enacting regulation, the UK govenment has thus set out a series of steps, which will include AML regulation pertaining to digital currency exchanges in the UK, to ensure that law enforcement bodies have the capabilities required to combat criminality in the digital currency space. More interventionist maybe than its European counterparts, the New York Department of Financial Services (NYDFS) has recently released the BitLicense regulatory framework, after approximately 2 years of consultation\footnote{The final text of the regulatory framework is available at \url{http://www.dfs.ny.gov/legal/regulations/adoptions/dfsp200t.pdf}.} The regulation sets out definitions for virtual currencies activities, which include:

\begin{itemize}
\item receiving virtual currency for transmission or transmitting virtual currency;
\item storing, holding, or maintaining custody or control of virtual currency on behalf of others; 
\item buying and selling virtual currency as a customer business; 
\item performing exchange services as a customer business; or 
\item controlling, administering, or issuing a virtual currency. 
\end{itemize}

Any individual or corporation engaged in the aforementioned activities is required to obtain a license to do so. This entails the completion of a lengthy application form\footnote{\url{http://www.dfs.ny.gov/legal/regulations/vc_license_application.pdf}} and a \$5000 fee. The regulation is far-reaching and there have already been firms that have either withdrawn their New York operations, or shut down altogether, citing excessive compliance burdens\footnote{\url{http://cointelegraph.com/news/114623/bitlicense-doing-its-job-eobot-becomes-3rd-firm-gone-from-new-york}}.

The Law Library of Congress has compiled a list of regulatory responses besides the ones detailed above \footnote{\url{http://www.loc.gov/law/help/bitcoin-survey/regulation-of-bitcoin.pdf}}. Outside of the EU and the US, regulatory activity regarding crypto-currency usage has mostly been limited to warning about its nature as a non-state-backed currency and its price volatility. There are a number of exceptions however, as China has banned financial institutions from handling bitcoin, while Japan has stated that `due to their intangible nature and reliance on third parties', bitcoins are effectively not subject to ownership, and thus are not covered by existing regulation\footnote{\url{http://www.japantimes.co.jp/news/2015/08/06/national/crime-legal/bitcoins-lost-in-mt-gox-debacle-not-subject-to-ownership-claims-tokyo-court-rules/\#.VctCRLwy3CK}}. On the other hand, the Australian Senate will effectively put forward recommendations to treat Bitcoin as money, as treating Bitcoin as a tradeable commodity would have created a double taxation effect\footnote{\url{http://www.reuters.com/article/2015/08/05/us-australia-bitcoin-idUSKCN0QA0TS20150805}}.    

A common theme in recent regulatory responses is that they have identified that more promising perspectives of virtual currencies may actually lie in the technology they use, i.e. the distributed ledger technologies introduced in Section \ref{sec:blockchain}. The term `virtual currency scheme' also encompasses the technologies and mechanisms used for the operation of transactions in the currency. The UK government, whilst identifying barriers that would prevent digital currencies from gaining widespread acceptance, has also identified the associated blockchain, or distributed ledger technology as having promise for the future of payments. Following the survey of HM Treasury, it has set out a series of recommendations to provide funding to research bodies to explore opportunities for digital currency technology. 

\section{Conclusions}
Our report highlights current trends in the virtual and crypto-currency space, from a number of different perspectives. The first is the emergence of such currencies, given the historical context of fiat money and the advent of cryptographic protocols that enabled e-money. We show that from this perspective, virtual currencies emerged to serve the need of particular niches of online gaming and social communities, while crypto-currencies sought to have a wider reach, and become the de facto currencies of the internet.   

Given these goals and the much greater probability for decentralised crypto-currencies to start entering the real economy, we focus on these to present current usage trends. Though to date, even the most popular crypto-currency, Bitcoin, has not gained widespread acceptance, while its use as an investment product has also remained low. It is believed that this will change as a greater understanding of these crypo-currencies occurs by regulators, exchanges and businesses in the economy. We hope to have contributed to this discussion by highlighting several aspects of monetary theory and the role of virtual and crypto-currencies in such theories.

Finally, we summarised current regulatory responses, showing the varied reaction to Bitcoin, from outright bans in China to effective treatment as money in Australia. The decentralised nature of the currency means that there is limited effect any single jurisdiction can have on the operation currency itself, and the focus is on companies providing services in the field. Given the borderless nature of Bitcoin, however, it is difficult to see how regulators can prevent companies taking advantage of regulatory arbitrage, by setting up in jurisdictions with less restrictions. 

\vspace{2em}
\textbf{Funding}

The support of the Economic and Social Research Council (ESRC) in funding the Systemic Risk Centre is gratefully acknowledged [grant number ES/K002309/1].

\bibliographystyle{authordate1}
\bibliography{bit}

\end{document}